\documentclass[10pt, conference, compsocconf]{IEEEtran}
\IEEEoverridecommandlockouts
\usepackage{cite}
\usepackage{amsmath,amssymb,amsfonts}
\usepackage{algorithm}
\usepackage{algpseudocode}
\usepackage{graphicx}
\usepackage{textcomp}
\usepackage{xcolor}
\usepackage{microtype}
\def\BibTeX{{\rm B\kern-.05em{\sc i\kern-.025em b}\kern-.08em
    T\kern-.1667em\lower.7ex\hbox{E}\kern-.125emX}}

\usepackage{subcaption}
\newcommand{\CX}{\ensuremath{\mathit{CX}}}

\begin{document}

\title{Scheduling of Operations in Quantum Compiler
$^*$
\thanks{*) © 2020 IEEE. Personal use of this material is permitted. Permission from IEEE must be obtained for all other uses, including reprinting/republishing this material for advertising or promotional purposes, collecting new collected works for resale or redistribution to servers or lists, or reuse of any copyrighted component of this work in other works.
Full citation of this paper: Toshinari Itoko and Takashi Imamichi. ``Scheduling of Operations in Quantum Compiler,'' in \textit{Proceedings of the International Conference on Quantum Computing and Engineering}. IEEE, 2020, pp. 337-344.}
}

\author{\IEEEauthorblockN{Toshinari Itoko}
\IEEEauthorblockA{\textit{IBM Quantum} \\
\textit{IBM Research - Tokyo}}
\and
\IEEEauthorblockN{Takashi Imamichi}
\IEEEauthorblockA{\textit{IBM Quantum} \\
\textit{IBM Research - Tokyo}}
}

\maketitle

\begin{abstract}
When scheduling quantum operations, a shorter overall execution time of the resulting schedule yields a better throughput and higher fidelity output.
In this paper, we demonstrate that quantum operation scheduling can be interpreted as a special type of job-shop problem.
On this basis, we provide its formulation as Constraint Programming while taking into account commutation between quantum operations.
We show that this formulation improves the overall execution time of the resulting schedules in practice
through experiments with a real quantum compiler and quantum circuits from two common benchmark sets.
\end{abstract}


\section{Introduction} \label{sec:introduction}

Although rapid progress in quantum computing device technology has
dramatically increased the coherence time of quantum bits (or
qubits), the currently available quantum computers remain in the
so-called noisy intermediate scale quantum regime~\cite{preskill2018quantum}.
For noisy quantum computers, it is important to schedule the operations on qubits
to be as short as possible because this increases the probability
of completing all of the operations before any qubit decoheres,
thus obtaining computational results with higher fidelity.
Even for fault-tolerant quantum computers,
shortening the duration of compiled schedules would increase the throughput.

Compilers for quantum computers (or quantum compilers) take a quantum circuit,
which is a sequence of quantum operations, as an input program
and generate a corresponding sequence of control instructions that are executable on the target hardware.
For example, in the case of quantum computers using superconducting qubits,
a quantum operation is compiled into several controls (e.g., a microwave pulse) for a certain period of time.
In general, any given quantum operation has its own processing time
and occupies its acting qubits for the duration as a computational resource.
For this reason, scheduling, through which the execution start time of each quantum operation is determined
without any overlapping, is an essential task in quantum compilers.
We call this task \emph{quantum operation scheduling}.
In this paper, we aim to minimize the overall execution time.
In the context of scheduling tasks across multiple resources
(qubits, in the case of quantum operation scheduling),
the time between the start of the first task and the end of the final task across all resources
is known as the makespan of the schedule.
Schedule length, overall execution time, and makespan are used interchangeably in this work.

The rough compilation flow considering the quantum operation scheduling task independently is as follows.
\begin{enumerate}
\item Gate decomposition: A task decomposes unitary operations called \emph{gates} with three or more qubits into those with one or two qubits.
\item Local simplification: A task simplifies a specific sequence of gates into one gate (or cancels them out).
\item Qubit routing: A task transforms a given circuit into an equivalent circuit so that all two-qubit gates are executed on limited pairs of qubits (depending on the physical implementation of the quantum computing device).
\item Quantum operation scheduling.
\item Control instruction mapping: A task maps each quantum operation to the corresponding control instructions.
\end{enumerate}

Most of the previous studies on quantum compilers have handled quantum operation scheduling within the context of its before and after tasks,
e.g., qubit routing and control instruction mapping
~\cite{venturelli2017temporal,metodi2006scheduling,guerreschi2018two,shi2019optimized}.
However, in practice, decomposing an entire compilation job into independent tasks
is becoming more common in the software architecture of quantum compilers,
similar to that of classical compilers, e.g.,~\cite{qiskit}.
Therefore, we focus on the following research question:
How much can we optimize the resulting schedule in quantum operation scheduling by itself?

In this paper, we examine quantum operation scheduling (QOS)
and analyze its theoretical properties and practical usefulness.
Our main contributions are as follows.
\begin{itemize}
\item
We show that QOS obtains greater degrees of freedom for optimizing the resulting schedule
by further considering the commutativity of particular quantum operations (Section~\ref{sec:qos}).
\item
We demonstrate that QOS can be reduced to a special type of job-shop problem
that has a disjunctive graph representation (Section~\ref{sec:qos-jsp})
so that we can formulate QOS as Constraint Programming and Mixed Integer
Programming, which are common techniques for the job-shop problem or scheduling in general
(Section~\ref{sec:solution}).
\item
We demonstrate through experiments with two common benchmark sets that the consideration of commutativity in QOS reduces
the schedule length by up to 7.36\%~(Section~\ref{sec:experiment}).
\end{itemize}

\section{Related Work} \label{sec:related-work}
The job-shop problem, also known as job shop scheduling, is a well known optimization problem in computer science and operations research, and many variations of it have been studied~\cite{blazewicz1996job, zhang2019review}.
See Section~\ref{sec:jsp} for its definition.
Algorithms to solve this problem include exact ones such as branch-and-bound based on a Mixed Integer Programming (MIP) formulation~\cite{manne1960job},
heuristic ones such as shifting bottleneck~\cite{adams1988shifting}, and
meta-heuristic ones such as simulated annealing~\cite{van1992job}.
In this paper, we mainly focus on the exact algorithm, as we want to determine the effect of optimizing quantum operation scheduling (QOS).

Task scheduling, which is the scheduling of computational tasks on multiple classical processors, has been extensively studied~\cite{topcuoglu2002performance,sinnen2007task}.
A special type of task scheduling, Directed Acyclic Graph (DAG) scheduling, which deals with heterogeneous processors~\cite{canon2008comparative,valouxis2013dag},
is most similar to QOS, but it differs in the way that resource constraints are handled.
In DAG scheduling, every task can be executed on any processor with a different cost, i.e., the resource constraint is soft, while in quantum operation scheduling, every quantum operation has fixed qubit operands that are not interchangeable, i.e., the resource constraint is hard.

Qubit routing is a task that transforms a given circuit into an equivalent circuit so that all two-qubit operations in it can be executed on limited pairs of qubits.
Schedule length can be approximated by circuit depth, which is the schedule length when assuming all operations have the same unit processing time.
Therefore, qubit routing with the objective of minimizing the circuit depth~\cite{maslov2008quantum,bhattacharjee2017depth} or two-qubit gate depth~\cite{cowtan2019qubit,childs2019circuit} can be viewed as
approximate quantum operation scheduling with qubit routing.
Although the algorithms for this may be applicable to scheduling without qubit routing,
they provide only approximate solutions, not the exact ones to QOS.

While we define quantum operation scheduling independently of hardware technology,
there are several studies on scheduling specialized for quantum computers based on ion trap technology~\cite{mohammadzadeh2009improving, bahreini2015minlp}.
These works consider a combination of scheduling and qubit routing
under a hardware structure model, called macroblocks,
and propose heuristic algorithms to solve it.

Several studies have considered the commutation of quantum operations in scheduling~\cite{venturelli2017temporal,metodi2006scheduling,guerreschi2018two,shi2019optimized}.
Venturelli et al.~\cite{venturelli2017temporal} examined the scheduling of quantum operations as a subproblem of qubit routing.
They proposed an exact method using a temporal planner and showed it works well for QAOA circuits, which have many commuting gates.
Although their method is applicable to quantum operation scheduling without qubit routing,
our methods discussed in Section~\ref{sec:solution} are simpler and perform sufficiently well for the specific scheduling problem considered in this paper.
Guerreschi and Park~\cite{guerreschi2018two} proposed a two-step solution that decomposes the problem with qubit routing and
solves quantum operation scheduling (without qubit routing) in the first step.
They provide a list scheduling heuristic algorithm using upward ranking but not any exact algorithm for scheduling.
Other studies have considered quantum operation scheduling
as a subtask of qubit routing~\cite{metodi2006scheduling}
or control optimization~\cite{shi2019optimized}.
While they provide practical heuristic algorithms for solving the task
that includes scheduling,
the exact algorithm for scheduling is not discussed.

\section{Problem} \label{sec:problem}

\subsection{Quantum Operation Scheduling} \label{sec:qos}
\begin{figure}[htbp]
\centerline{\includegraphics[clip, width=.25\textwidth]{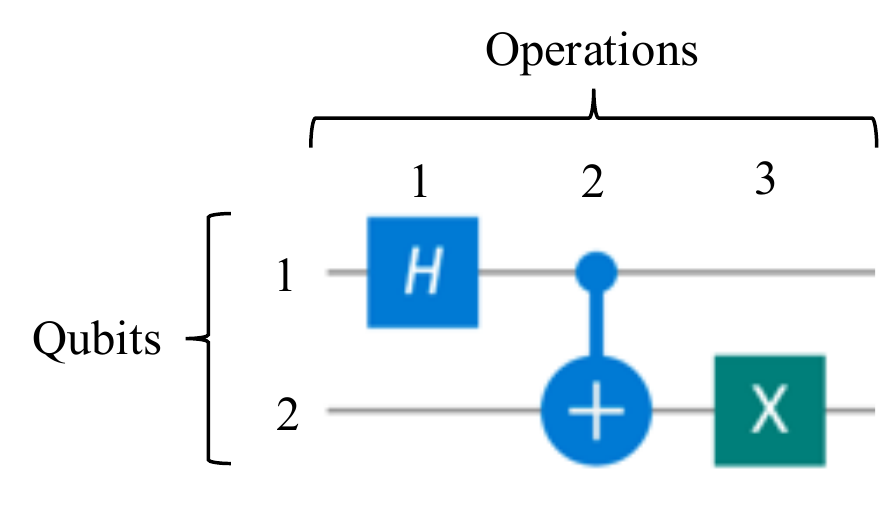}}
\caption{Diagram representation of a quantum circuit}
\label{fig:qc}
\end{figure}

\begin{figure}[htbp]
  \centering
  \begin{subfigure}[b]{0.17\textwidth}
      \includegraphics[clip, width=\textwidth]{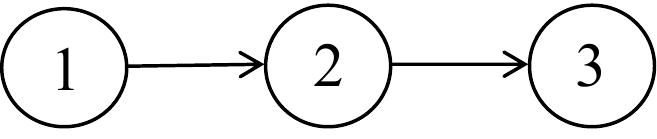}
      \medskip
      \caption{Standard DAG}\label{fig:dag-std}
  \end{subfigure}
  ~
  \begin{subfigure}[b]{0.27\textwidth}
      \includegraphics[clip, width=\textwidth]{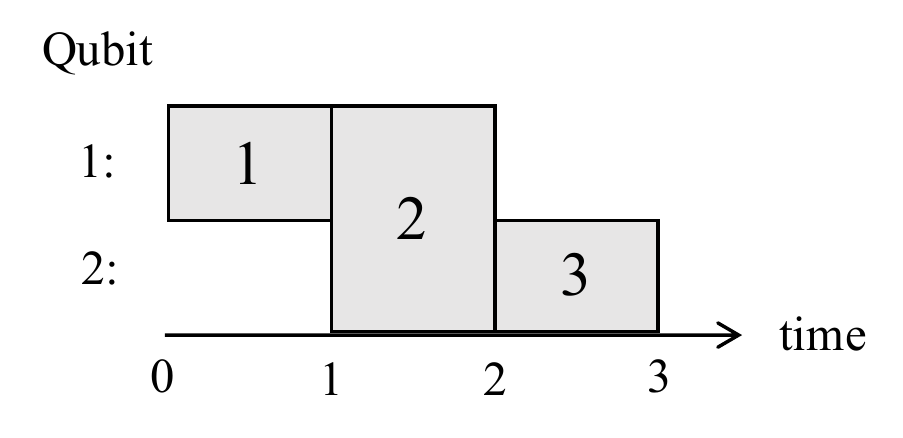}
      \caption{Schedule with makespan = 3}\label{fig:sched-3}
  \end{subfigure}
  \caption{Standard dependency graph and resulting schedule}\label{fig:std-sched}
\end{figure}

\begin{figure}[htbp]
  \centering
  \begin{subfigure}[b]{0.17\textwidth}
      \centering
      \includegraphics[clip, width=.6\textwidth]{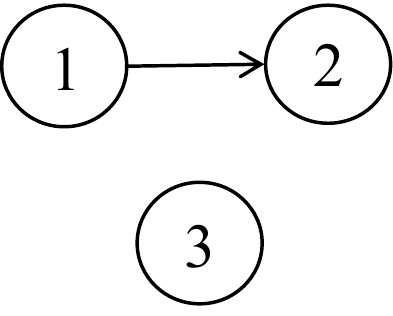}
      \smallskip
      \caption{Extended DAG}\label{fig:dag-ext}
  \end{subfigure}
  ~
  \begin{subfigure}[b]{0.27\textwidth}
      \includegraphics[clip, width=\textwidth]{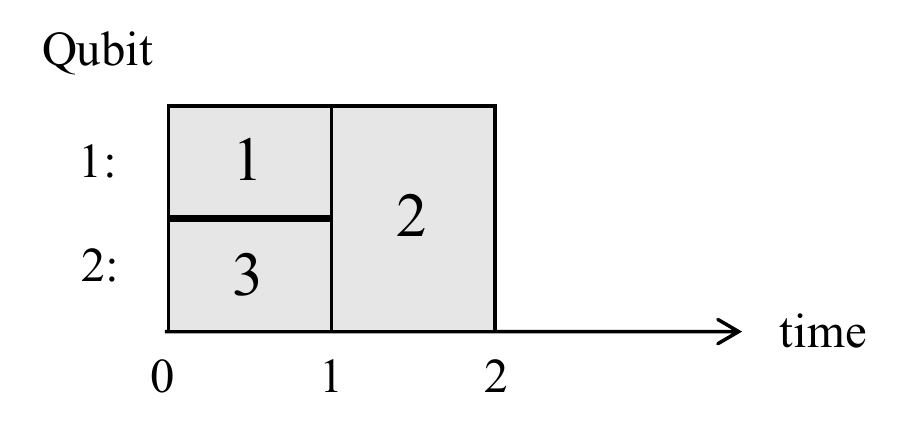}
      \caption{Schedule with makespan = 2}\label{fig:sched-2}
  \end{subfigure}
  \caption{Extended dependency graph and resulting schedule}\label{fig:ext-sched}
\end{figure}

We define quantum operation scheduling as
the problem of finding a \emph{schedule} for a given quantum circuit.
A quantum circuit is a sequence of quantum operations.
Many of them are unitary operations called \emph{gates}.
Each of the quantum operations has acting qubits and its own processing time.
A quantum circuit is given as a sequence: e.g., $[H(1), \CX(1, 2), X(2)]$.
Here, $H(1)$ denotes a Hadamard gate acting on qubit $1$,
$\CX(1, 2)$ denotes a Controlled-NOT (or CNOT) gate acting on control qubit $1$ and target qubit $2$,
and $X(2)$ denotes a NOT gate acting on qubit $2$.
Quantum circuits are typically depicted in a circuit diagram, as shown in Fig.~\ref{fig:qc}.
For simplicity, we assume all of the operations have the same unit processing time in Fig.~\ref{fig:qc}.
If we ignore commutation between gates,
the gate dependency graph is linear,
i.e., $H(1)$ must precede $\CX(1, 2)$ and $\CX(1, 2)$ must precede $X(2)$,
and we obtain a trivial schedule (makespan = 3), as shown in Fig.~\ref{fig:std-sched}.
We call the graph representing the dependencies among gates in a circuit the \emph{dependency graph}.
In contrast, if we consider that $\CX(1, 2)$ and $X(2)$ commute,
we have a different dependency graph: $H(1)$ must precede $\CX(1, 2)$, but there is no restriction on $X(2)$, so
we can obtain a shorter schedule (makespan = 2), as shown in Fig.~\ref{fig:ext-sched}.
This is compelling evidence that commutation rules should be considered when scheduling circuit operations.
%

A schedule is defined by the start times of the operations in a given circuit.
Any schedule must satisfy two elementary constraints:
\emph{precedence} and \emph{non-overlap}.
The precedence constraint restricts the execution order of operations to obey a partial order represented as a dependency graph.
The non-overlap constraint allows only the processing of one operation on a qubit at a time.

Generally, the supported basic operations and the processing times depend on the target hardware.
Hereafter, we assume that basic operations are given and that all circuits have already been decomposed into them.
We also assume that each processing time of the basic operations is fixed and given as a parameter.
The dependency graph of a provided quantum circuit
varies depending on which commutation rules are considered.
Taking these details into account, we formally define a quantum operation scheduling problem as follows.
\newtheorem{qos}{Quantum Operation Scheduling}
\renewcommand{\theqos}{}
\begin{qos}
Given a quantum circuit as a sequence of basic operations,
each processing time of each operation, and 
a set of commutation rules between basic operations,
find a schedule that satisfies precedence and non-overlap constraints with the minimum makespan.
\end{qos}

\subsection{Job-shop Problem and its Disjunctive Graph Representation} \label{sec:jsp}
We review a basic version of the job-shop problem as follows.
Let $J=\{J_1,\ldots,J_n\}$ be a set of $n$ jobs and
$M=\{M_1,\ldots,M_m\}$ be a set of $m$ machines.
Each job $J_j$ has an operation sequence $O_j$ to be processed in a specific order,
called the \emph{precedence constraint}.
We denote the $k$-th operation in $O_j$ by $O_{jk}$.
Each operation $O_{jk}$ requires exclusive use of a specific machine for its processing time $p_{jk}$,
called the \emph{non-overlap constraint}.
A schedule is a set of start (or completion) times for each operation $t_{jk}$ that satisfies both constraints.
The objective of the job-shop problem is minimization of the makespan.

\begin{figure}[htbp]
\centerline{\includegraphics[clip, width=.40\textwidth]{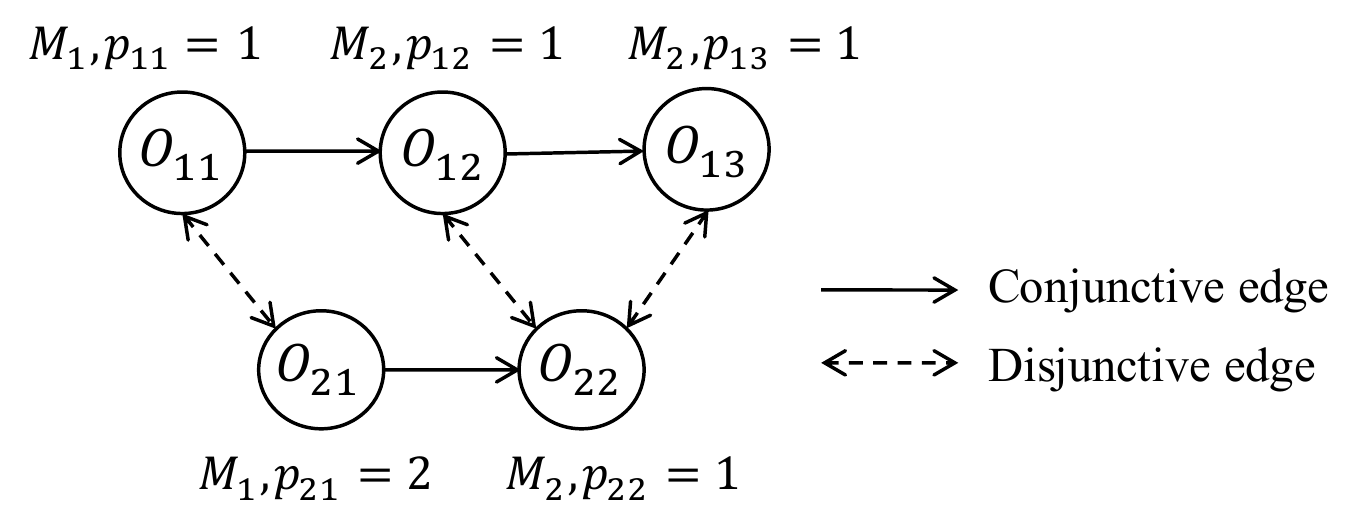}}
\caption{Disjunctive graph representation of a job-shop problem}
\label{fig:jsp-graph}
\end{figure}

The job-shop problem is often represented by a disjunctive graph $G = (V, C \cup D)$,
where
\begin{itemize}
\item $V$ is a set of nodes representing the operations $O_{jk}$,
\item $C$ is a set of conjunctive (directed) edges representing the order of the
operations in any job, and
\item $D$ is a set of disjunctive edges representing pairs of operations that must be processed on the same machine.
\end{itemize}
For each node, the processing time and the required machine of its corresponding operation is attached.
Conjunctive edges $C$ represent the precedence constraint and disjunctive edges $D$ represent the non-overlap constraint.
Note that disjunctive edges whose direction is fixed by some conjunctive edges can be omitted.
That means any disjunctive edge can be removed if there exists a path from one end of the edge to the other on a conjunctive graph $(V, C)$.
Figure~\ref{fig:jsp-graph} shows an example of a disjunctive graph representing the job-shop problem.
The operation $O_{11}$ must be processed in machine $M_1$ and it takes $1$ time unit.
The disjunctive edge $(O_{12}, O_{13})$ is omitted
since its direction is fixed by the conjunctive edge at the same place.

On the basis of this disjunctive graph representation,
the job-shop problem can be seen as a problem of determining the direction of disjunctive edges while keeping the resulting graph acyclic.
This is equivalent to determining the ordering of the operations processed on the same machine,
and such ordering yields a unique schedule,
called a \emph{semi-active schedule}~\cite{fleming1997genetic},
by sequencing operations as early as possible.
\begin{figure}[htbp]
\centerline{\includegraphics[clip, width=.50\textwidth]{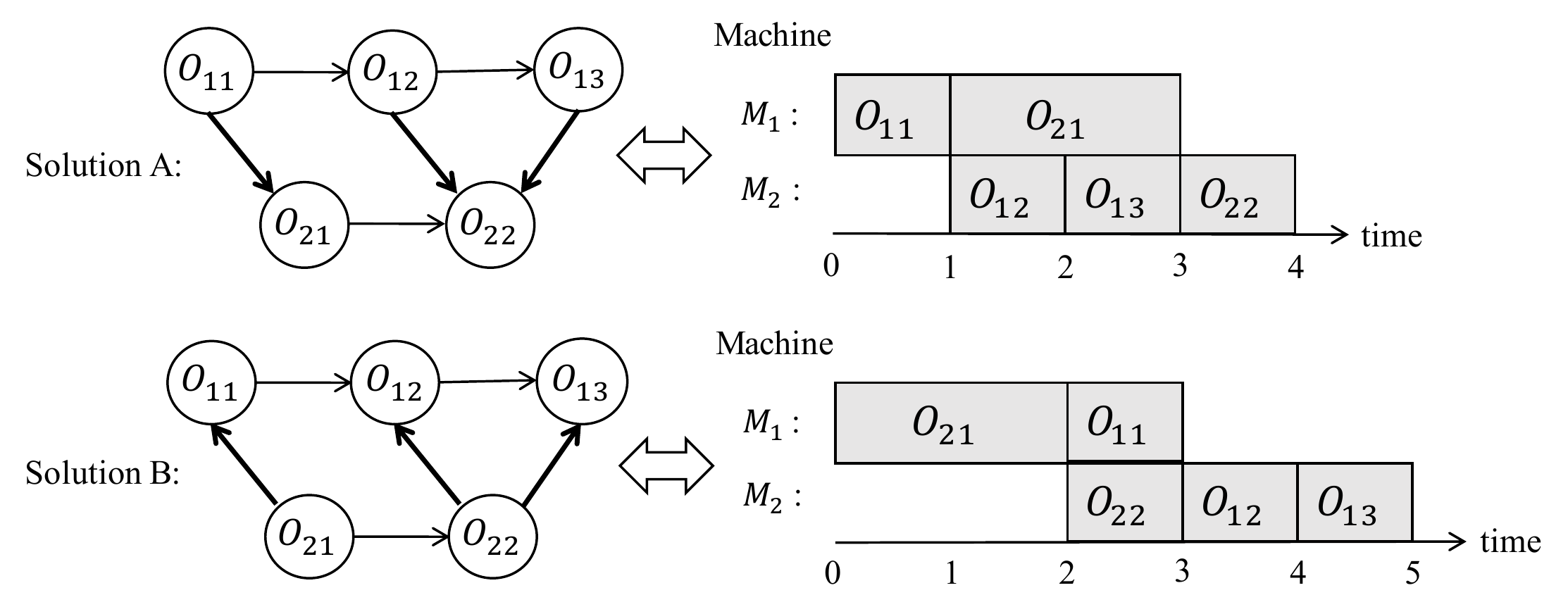}}
\caption{Two solutions to the job-shop problem in Fig.~\ref{fig:jsp-graph}}
\label{fig:jsp-solutions}
\end{figure}
Figure~\ref{fig:jsp-solutions} shows two solutions to the job-shop problem
defined by the disjunctive graph depicted in Fig.~\ref{fig:jsp-graph}.
As shown, a different selection of the direction of the disjunctive edges results in
a different solution.
Among the operations $\{O_{12}, O_{13}, O_{22}\}$,
we cannot select \{$(O_{22}, O_{12})$, $(O_{13}, O_{22})$\}
because it produces a cycle $O_{12} \rightarrow O_{13} \rightarrow O_{22} \rightarrow  O_{12}$.
In Solution A, the directed edge $(O_{11}, O_{21})$ determines the order of operations processed on machine $M_1$,
and \{$(O_{12}, O_{22})$, $(O_{13}, O_{22})$\} determine that on machine $M_2$.

\subsection{Disjunctive Graph Representation of Quantum Operation Scheduling} \label{sec:qos-jsp}
Technically, quantum operation scheduling can be seen as a special type of job-shop problem with the following properties:
(1) one job has one operation,
(2) a precedence constraint is given as a partial ordering among all operations
instead of total ordering of operations per job, and
(3) multiple machines (i.e., qubits) can be occupied by a single operation at the same time.
Those properties preserve enough conditions to represent the problem by a disjunctive graph $G = (V, C \cup D)$.
For property (1), we need to define the problem on operations without jobs,
but this does not change the fact that nodes $V$ represent operations.
For properties (2) and (3), we need to modify the definition of conjunctive edges $C$
and disjunctive edges $D$, respectively as follows.

The quantum operation scheduling can be represented by a disjunctive graph $G = (V, C \cup D)$, where
\begin{itemize}
\item $V$ is a set of nodes representing the quantum operations in a given circuit,
\item $C$ is a set of conjunctive edges representing dependencies among the operations, i.e., edges of the dependency graph, and
\item $D$ is a set of disjunctive edges representing pairs of operations that act on the same qubit and (possibly) commute with one another.
\end{itemize}
For each operation (i.e., node) $i \in V$,
the processing time $p_i$ and the acting qubits are attached.
Note that conjunctive graph $(V, C)$ is a DAG given as a dependency graph.
Conjunctive edges $C$ and disjunctive edges $D$ still represent the precedence constraint and non-overlap constraint, respectively.
It is known that the dependency graph for any circuit can be computationally constructed under several popular commutation rules~\cite{itoko2020optimization}.
When provided with the dependency graph of a quantum circuit,
the disjunctive graph representation can be computationally constructed.
In fact, it is possible to define disjunctive edges by all the pairs of nodes that acting on the same qubit.
This definition is redundant, but there is no problem with including edges (pairs of operations) that do not commute with one another.
The point is that it must include all of the commuting pairs.
Starting from the redundant edges,
we can define the minimal disjunctive edges
by removing edges that have a path on the conjunctive graph.
However, this comes at a significant computation cost.
There is an in-between definition that
picks up operations acting on any qubit and
splits them into sets of operations commuting each other within each of the sets.
We used this definition for the experiments discussed in Section~\ref{sec:experiment}.
Although this cannot provide the minimal edges
because the operations within a set may not commute each other
when considering operations acting on the other qubits,
it has fewer edges than the most redundant definition
and requires less computation cost than the minimal definition.

With the disjunctive graph representation of the quantum operation scheduling,
we take the commutation of operations into account on the basis of the difference of the dependency graph.
We call the dependency graph that considers only the trivial commutation between operations not sharing their acting qubits \emph{standard DAG}
and the dependency graph that considers commutation rules in addition to the trivial ones \emph{extended DAG}.

\begin{figure}[htbp]
  \centering
  \begin{subfigure}[b]{0.24\textwidth}
      \includegraphics[clip, width=\textwidth]{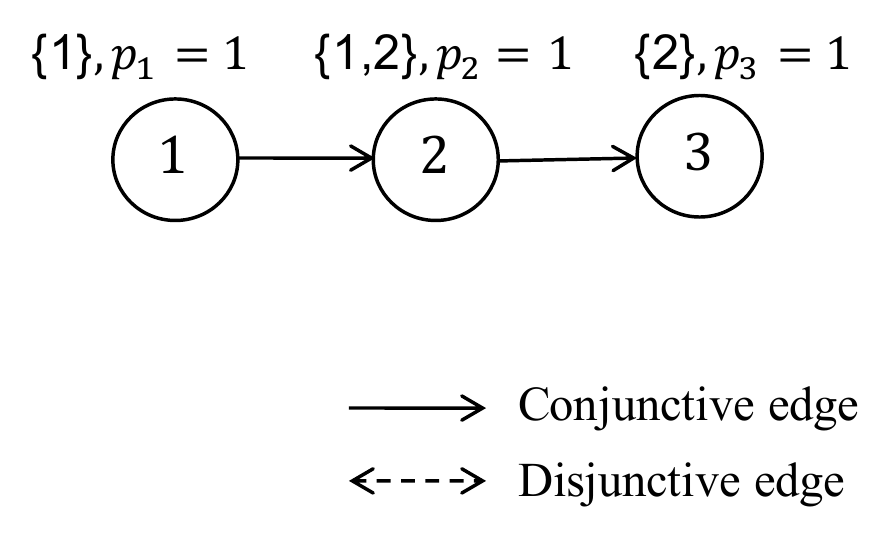}
      \caption{Standard DAG}\label{fig:qos-std}
  \end{subfigure}
  \begin{subfigure}[b]{0.17\textwidth}
      \includegraphics[clip, width=\textwidth]{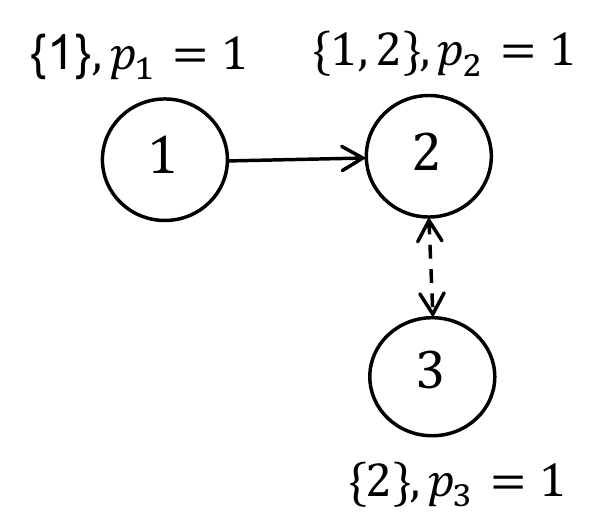}
      \caption{Extended DAG}\label{fig:qos-ext}
  \end{subfigure}
  \caption{Disjunctive graphs representing quantum operation scheduling for different dependency graphs}\label{fig:qos-graphs}
\end{figure}

Figure~\ref{fig:qos-graphs} shows disjunctive graphs
representing two quantum operation scheduling problems
that have the same operations but different dependency (conjunctive) graphs
(standard DAG and extended DAG).
In the case of extended DAG, we can select the order of $O_2$ and $O_3$,
while in standard DAG, all the orders among operations are fixed.

It generally holds that if
we consider a standard DAG, there are no disjunctive edges, i.e., $D=\emptyset$.
This means there is a unique semi-active schedule
because each ordering of the operations processed on the same qubit is uniquely determined by the conjunctive DAG.
However, if we consider an extended DAG,
we have fewer edges in $C$ and some edges in $D$.
This creates the room for selecting a better ordering of the operations processed on the same qubit, i.e., optimizing the schedule.

\section{Formulation} \label{sec:solution}
We provide a Constraint Programming (CP) formulation and a Mixed Integer Programming (MIP) formulation for the quantum operation scheduling problem (QOS) defined in the previous section.
By solving them, we can find the optimal solution of QOS and
analyze how much we can improve the resulting schedule in QOS.
In this section, we assume that the disjunctive graph representation of quantum operation scheduling $G = (V, C \cup D)$ for a given circuit has already been constructed.

\subsection{Constraint Programming Formulation} \label{sec:cp}
Let $x_i$ be an interval variable describing the start and end time of operation $i \in V$,
whose duration is fixed to its processing time $p_i$.
Using functions commonly supported by CP solvers, e.g. $\mathbf{interval\_var}$,
quantum operation scheduling is formulated as a CP:
  \[
    \begin{array}{ll}
  \mbox{minimize} & \max \{\mathbf{end\_of}(x_i) \mid i \in V\} \\
  \mbox{subject to} & \mathbf{end\_before\_start}(x_i, x_j),\ \forall (i, j) \in C,\\
        & \mathbf{no\_overlap}(x_k, x_l),\ \forall (k, l) \in D,\\
        & x_i \equiv \mathbf{interval\_var}(\mathrm{duration}=p_i),\ \forall i \in V.\\
    \end{array}
  \]
Here $\mathbf{end\_of}(x_i)$ takes the end time of $x_i$, so the objective is the minimization of the makespan.
The constraint $\mathbf{end\_before\_start}(x_i, x_j)$ means
the end time of $x_i$ must precede the start time $x_j$, so it represents the precedence constraint.
The constraint $\mathbf{no\_overlap}(x_k, x_l)$ means
the interval $x_k$ must not overlap the interval $x_l$, so it represents the non-overlap constraint.
Note that, for any operation pair $(i, j) \not\in D$, the precedence constraint guarantees no overlap between them.
All of $\mathbf{end\_of}$, $\mathbf{end\_before\_start}$, $\mathbf{no\_overlap}$, and $\mathbf{interval\_var}$ are supported in the IBM ILOG CP Optimizer.

\subsection{Mixed Integer Programming Formulation} \label{sec:mip}
Let $x_i$ be a variable representing the start time of operation $i \in V$ and
$p_i$ be a constant parameter representing its processing time.
Let $t$ be a makespan of the schedule defined by $x$.
Let $y_{kl}$ be an indicator (Boolean) variable that takes True (1) if operation $k$ precedes operation $l$ and False (0) if $l$ precedes $k$.
Using these, quantum operation scheduling is formulated as a MIP:
  \[
    \begin{array}{ll}
  \mbox{minimize} & t \\
  \mbox{subject to} & x_i + p_i \leq x_j,\ \forall (i, j) \in C,\\
        & \ y_{kl} \Rightarrow x_k + p_k \leq x_l,\ \forall (k, l) \in D,\\
        & \neg y_{kl} \Rightarrow x_l + p_l \leq x_k,\ \forall (k, l) \in D,\\
        & x_i + p_i \leq t,\ \forall i \in V,\\
        & 0 \leq x_i \in \mathbb{R},\ \forall i \in V,\\
        & y_{kl} \in \{0, 1\},\ \forall (k, l) \in D.\  \\
    \end{array}
  \]
The inequality $x_i + p_i \leq x_j$ represents the precedence constraint.
The two inequalities using $\Rightarrow$ represent the non-overlap constraint.
Note that the constraints with indicator variables $y$ can be translated into linear constraints by applying the so-called the big-M technique.
However, recent MIP solvers have the capability to handle such indicator constraints very well, so we leave the formulation with indicator variables.

\section{Experiment} \label{sec:experiment}

We conducted two experiments.
In the first one, we evaluate how much the consideration of commutation between operations improves the schedule in quantum operation scheduling.
In the second one, we investigate how much the extent of improvement in scheduling can be affected by the optimization level in a previous task.

\subsection{Common Experimental Settings}
Both of the experiments were conducted in a real compiling environment.
As the target quantum computing device for compilation,
we used the IBM Q Johannesburg, which has 20 qubits (see~\cite{corcoles2019challenges} for the details).
We implemented our scheduling algorithms within Qiskit 0.18.0 (Terra 0.13.0),
which is an open-source quantum computing software development framework~\cite{qiskit}.

We first transpiled all of the circuits to make them executable on the IBM Q Johannesburg, i.e., we solved the qubit routing problem to map given circuits onto the device topology.
For this, we used the ``transpile()'' function in Qiskit and set the ibmq\_johannesburg backend, fixed the seed\_transpiler to 1, the optimization\_level to 2, and left the other options as the default.
During transpiling, all of the circuits were decomposed into the basis gates $\{u1, u2, u3, \CX\}$, which are elementary gates supported by the backend.
Here $u1, u2, u3$ are single-qubit gates and $\CX$ is a two-qubit gate.
The execution time for each gate (gate length) is provided as the backend properties.
Note that it can differ depending on which qubit(s) the gate acts on,
e.g. the length of $u3(1)$ can be different from that of $u3(2)$.
We used those as of April 11, 2020.

We then applied our scheduling algorithms to the transpiled circuits.
We used the real processing time for each of the basis gates provided as backend properties.
We considered three commutation rules on the basis gates---$u1(i)\leftrightarrow\CX(i,j)$,
$\CX(i,j)\leftrightarrow\CX(i,k)$, and
$\CX(i,k)\leftrightarrow\CX(j,k)$---in the construction of the extended DAGs for the transpiled circuits.

We used the IBM ILOG CP Optimizer and CPLEX 12.9.0
to solve the scheduling problem based on the CP and MIP formulation described in Section~\ref{sec:solution}, respectively.

\subsection{Improvement by Considering Commutation in Quantum Operation Scheduling} \label{sec:experiment-improve}
In the first experiment, we quantified the significance of considering the commutation of operations in quantum operation scheduling.
Specifically, we evaluated the improvement by comparing the best solutions (makespans) of the formulation constrained by standard DAG with those by extended DAG as shown in Table~\ref{tab:improvement}.

\begin{table}[htbp]
 \centering
 \caption{Comparison of makespans [$dt$] ($1\,dt=2/9\,$ns) obtained by the formulation based on standard-DAG (Std-DAG) and those based on extended DAG (Ext-DAG) for 16 circuits from the RevLib benchmark.
 For the Ext-DAG formulation, the solutions by the Constraint Programming solver with the time limit of ten seconds are listed.
 The Qubits and Gates columns list the number of qubits and gates in the input circuits.
 The $\Delta$ column lists the improvement rate from Std-DAG to Ext-DAG.}
 \label{tab:improvement}
\begin{tabular}{lrr|rrr}
\hline
Circuit name & Qubits & Gates & Std-DAG & Ext-DAG & $\Delta$ \\
\hline
mini\_alu\_305	 & 	10	 & 	173	 & 	24,940 	 & 	24,308 	 & 	2.53\%	 \\
qft\_10	 & 	10	 & 	200	 & 	24,358 	 & 	24,074 	 & 	1.17\%	 \\
sys6-v0\_111	 & 	10	 & 	215	 & 	30,604 	 & 	30,114 	 & 	1.60\%	 \\
rd73\_140	 & 	10	 & 	230	 & 	35,848 	 & 	35,484 	 & 	1.02\%	 \\
ising\_model\_10	 & 	10	 & 	480	 & 	4,210 	 & 	4,210 	 & 	0.00\%	 \\
wim\_266	 & 	11	 & 	986	 & 	141,914 	 & 	138,658 	 & 	2.29\%	 \\
sym9\_146	 & 	12	 & 	328	 & 	50,458 	 & 	50,170 	 & 	0.57\%	 \\
rd53\_311	 & 	13	 & 	275	 & 	40,420 	 & 	40,164 	 & 	0.63\%	 \\
ising\_model\_13	 & 	13	 & 	633	 & 	4,210 	 & 	4,210 	 & 	0.00\%	 \\
0410184\_169	 & 	14	 & 	211	 & 	40,356 	 & 	39,074 	 & 	3.18\%	 \\
sym6\_316	 & 	14	 & 	270	 & 	47,178 	 & 	46,404 	 & 	1.64\%	 \\
rd84\_142	 & 	15	 & 	343	 & 	39,812 	 & 	38,490 	 & 	3.32\%	 \\
cnt3-5\_179	 & 	16	 & 	175	 & 	19,630 	 & 	19,366 	 & 	1.34\%	 \\
cnt3-5\_180	 & 	16	 & 	485	 & 	69,854 	 & 	68,326 	 & 	2.19\%	 \\
qft\_16	 & 	16	 & 	512	 & 	50,674 	 & 	50,088 	 & 	1.16\%	 \\
ising\_model\_16	 & 	16	 & 	786	 & 	4,370 	 & 	4,370 	 & 	0.00\%	 \\  \hline
\end{tabular}
\end{table}

For this experiment, we used quantum circuits from the test dataset provided by Zulehner et al.~\cite{zulehner2018efficient}, which originated from the RevLib benchmark ~\cite{soeken2011revkit}.
We selected 16 circuits with 10--16 qubits and less than 1000 gates from among them.
We used the as-soon-as-possible heuristic scheduling algorithm implemented in Qiskit to find the unique solutions from the standard DAG formulation (Std-DAG).
We also used the CP solver with a 10-sec time limit to find the best possible solutions from the extended DAG formulation (Ext-DAG).

Looking at the $\Delta$ column in Table~\ref{tab:improvement}, i.e., the improvement rates from Std-DAG to Ext-DAG, we can see they were non-negative and varied depending on the circuit structures from 0.00\% to 3.32\% (median 1.26\%).
These results demonstrate that the commutation-aware formulation we proposed in Section~\ref{sec:qos} can improve the resulting schedule length in a practical situation.
Although they may look marginal, they may be welcomed by those who have abundant time for compilation and need more optimization.

\begin{table*}[htbp]
\centering
\caption{Difference in improvement rates ($\Delta$ column) of makespans from scheduling with standard DAG (Std-DAG) compared to those with extended DAG (Ext-DAG) using CP solver after applying a naive gate decomposition or optimized gate decomposition. }
\label{tab:diff}
\begin{tabular}{l|rrr|rrr}
\hline
Circuit name	 & 	\multicolumn{3}{|c|}{Naive gate decomposition}					 & 	\multicolumn{3}{c}{Optimized gate decomposition}					 \\
	 & 	Std-DAG	 & 	Ext-DAG	 & 	$\Delta$	 & 	Std-DAG	 & 	Ext-DAG	 & 	$\Delta$	 \\
\hline
Mod 5\_4	 & 	6,328 	 & 	5,984 	 & 	5.44\%	 & 	6,016 	 & 	5,578 	 & 	7.28\%	 \\
VBE-Adder\_3	 & 	19,126 	 & 	18,852 	 & 	1.43\%	 & 	11,416 	 & 	11,148 	 & 	2.35\%	 \\
CSLA-MUX\_3	 & 	22,238 	 & 	21,438 	 & 	3.60\%	 & 	19,774 	 & 	18,778 	 & 	5.04\%	 \\
RC-Adder\_6	 & 	28,606 	 & 	27,564 	 & 	3.64\%	 & 	20,408 	 & 	18,906 	 & 	7.36\%	 \\
Mod-Red\_{21}	 & 	33,160 	 & 	32,348 	 & 	2.45\%	 & 	28,320 	 & 	27,494 	 & 	2.92\%	 \\
Mod-Mult\_{55}	 & 	13,942 	 & 	13,740 	 & 	1.45\%	 & 	14,478 	 & 	14,070 	 & 	2.82\%	 \\
Toff-Barenco\_3	 & 	12,388 	 & 	12,388 	 & 	0.00\%	 & 	4,812 	 & 	4,812 	 & 	0.00\%	 \\
Toff-NC\_3	 & 	9,108 	 & 	9,108 	 & 	0.00\%	 & 	3,818 	 & 	3,818 	 & 	0.00\%	 \\
Toff-Barenco\_4	 & 	15,992 	 & 	15,582 	 & 	2.56\%	 & 	9,006 	 & 	8,678 	 & 	3.64\%	 \\
Toff-NC\_4	 & 	12,272 	 & 	11,834 	 & 	3.57\%	 & 	8,016 	 & 	7,626 	 & 	4.87\%	 \\
Toff-Barenco\_5	 & 	21,600 	 & 	21,378 	 & 	1.03\%	 & 	19,424 	 & 	19,138 	 & 	1.47\%	 \\
Toff-NC\_5	 & 	13,536 	 & 	12,984 	 & 	4.08\%	 & 	9,262 	 & 	8,920 	 & 	3.69\%	 \\
Toff-Barenco\_{10}	 & 	90,940 	 & 	89,248 	 & 	1.86\%	 & 	59,282 	 & 	57,282 	 & 	3.37\%	 \\
Toff-NC\_{10}	 & 	43,156 	 & 	42,546 	 & 	1.41\%	 & 	28,720 	 & 	28,322 	 & 	1.39\%	 \\
GF($2^4$)-Mult	 & 	33,160 	 & 	32,492 	 & 	2.01\%	 & 	33,646 	 & 	33,148 	 & 	1.48\%	 \\
GF($2^5$)-Mult	 & 	37,948 	 & 	36,480 	 & 	3.87\%	 & 	44,702 	 & 	44,030 	 & 	1.50\%	 \\
GF($2^6$)-Mult	 & 	63,856 	 & 	61,870 	 & 	3.11\%	 & 	64,784 	 & 	64,180 	 & 	0.93\%	 \\
\hline
\end{tabular}
\end{table*}

We also examined the MIP solver (with the same 10-sec time limit) to find solutions of the Ext-DAG; however, all of the solutions were slightly worse or equal to those by the CP solver.
Hence, we omitted these results in Table~\ref{tab:improvement}.
As for the solutions (i.e., makespans) from the standard DAG formulation,
we verified that those by the CP and MIP solvers were exactly the same as those by the as-soon-as-possible heuristic scheduling algorithm implemented in Qiskit as expected.

\subsection{Performance Variation by Optimization Level of Previous Task}

In the second experiment, we investigated how a previous task affects the solution quality in the quantum operation scheduling task.
To this end, as a previous task, we picked the gate decomposition task that decomposes gates with three or more qubits into those with one or two qubits.
We changed the optimization level in the gate decomposition task and observed how it affects the improvement rates of makespans from scheduling with standard DAG (Std-DAG) compared to those with extended DAG (Ext-DAG), as shown in Table~\ref{tab:diff}.

For this experiment, we used 17 circuits from the test dataset provided by Nam et al.~\cite{nam2018automated}.
To change the optimization level in the gate decomposition task,
we used both their input circuit data (with ``\_before'' suffix in their file names) and output circuit data after the heavy optimization proposed in~\cite{nam2018automated} (with ``\_after\_heavy'' suffix) as our input circuits to be scheduled.
Those correspond with the Naive gate decomposition column and Optimized gate decomposition column, respectively.
Note that, in the Naive case, gates are decomposed by a simple rule-based algorithm implemented in Qiskit before scheduling.
As in the previous experiment, we used the as-soon-as-possible heuristic scheduling algorithm implemented in Qiskit to find the unique solutions from the Std-DAG formulation
and the CP solver with a 10-sec time limit to find the best possible solutions from the Ext-DAG formulation.

As shown in the two $\Delta$ columns in Table~\ref{tab:diff},
we can observe clear improvement from Std-DAG to Ext-DAG
no matter which gate decomposition algorithm we used before QOS:
Min: 0.00\%--Median: 2.45\%--Max: 5.44\% (Naive) and
Min: 0.00\%--Median: 2.82\%--Max: 7.36\% (Optimized).
This again confirms that our commutation-aware formulation proposed in Section~\ref{sec:qos} can improve the resulting schedule length in a practical situation.

When we compare the improvement rates ($\Delta$ column) from scheduling after the naive gate decomposition with those after the optimized gate decomposition in Table~\ref{tab:diff},
there are two key findings.
First, they have a similar median: 2.45\% (Naive) and 2.82\% (Optimized).
This suggests that, on average, our commutation-aware scheduling can stably improve the resulting schedule no matter how much circuits has been optimized in a previous task (at least in the gate optimization task).
Second, the improvement rates for each individual circuit differs between Naive and Optimized.
Specifically, they increase from Naive to Optimized for ten circuits and decrease for five circuits.
This suggests that the optimization level of the previous task significantly affects the optimization gain in QOS.

Comparing the makespans in the Std- or Ext-DAG column under naive gate decomposition with those under optimized gate decomposition in Table~\ref{tab:diff},
we can see that they decrease for 13 out of 17 circuits, as expected, but increase for four circuits.
The latter four exceptional cases stem from negative interference among optimization tasks before scheduling, i.e., between gate decomposition and some tasks done within `transpile()` in Qiskit, and they are not caused by any errors in the scheduling.

\section{Discussion}
The basic version of the job-shop problem as a decision problem is known to be NP-complete~\cite{garey2002computers}.
Since QOS is a special variant of the job-shop problem, as discussed in Section~\ref{sec:qos-jsp}, it is not necessary for it to be NP-complete.
Identifying the theoretical complexity of QOS would be an interesting avenue for future work.

Throughout this paper, we have investigated how to minimize the overall execution time of the resulting schedule.
Although this certainly contributes to obtaining computational results with higher fidelity,
there should be more direct approaches that attempt to maximize the output fidelity
by considering gate-dependent errors.
In fact, such approaches have recently proposed in qubit routing~\cite{tannu2019not,murali2019noise,nishio2020extracting}.
Utilizing techniques like this for QOS is also left for future work.

The two formulations (CP/MIP) discussed in Section~\ref{sec:solution} are useful for the theoretical best case analysis because their solvers implement exact algorithms
that can find the optimal solution in the long run.
They may also be sufficient for certain practical applications, since CP/MIP solvers usually implement problem-agnostic heuristic algorithms to find the best possible solution within a limited time.
However, for the use cases where the compilation time is too critical to use CP/MIP solvers, it is worth considering heuristic algorithms specialized for QOS.
As an example, we developed a heuristic algorithm based on the Heterogeneous-Earliest-Finish-Time (HEFT) algorithm for task scheduling.
We provide its details in Appendix.
Such a heuristic algorithm can complement the CP/MIP-based approach.

\section{Conclusion}
We investigated quantum operations scheduling for the problem of scheduling quantum operations in a given circuit with the shortest total execution time.
We demonstrated that quantum operations scheduling can be interpreted as a special type of job-shop problem where we consider the commutation between quantum operations to make room for optimization.
We provided a Constraint Programming formulation and showed through experiments with real circuits and a compiler that
solving quantum operations scheduling independently improved the schedule length by the modest rate up to 7.36\%.






\section*{Acknowledgment}
We thank Dmitri Maslov, Lauren Capelluto, Thomas A. Alexander, and Rudy Raymond for their helpful comments.


\bibliographystyle{IEEEtran}
\bibliography{qc-scheduler}


\begin{table*}[htbp]
\centering
\caption{Makespans and their improvement rates ($\Delta$) from scheduling with standard DAG (Std-DAG) compared to those with extended DAG (Ext-DAG) using HEFT heuristic algorithm or CP solver after applying a naive gate decomposition or optimized gate decomposition. }
\label{tab:heft}
\begin{tabular}{l|rrr|rrr}
\hline
Circuit name	 & 	\multicolumn{3}{|c|}{Naive gate decomposition}
					 & 	\multicolumn{3}{c}{Optimized gate decomposition}	\\
	 & 	Std-DAG	 & 	\multicolumn{2}{c|}{Ext-DAG}
	 & 	Std-DAG	 & 	\multicolumn{2}{c}{Ext-DAG}	 \\
	 & 	 & 	\multicolumn{1}{c}{HEFT ($\Delta$) }	& 	\multicolumn{1}{c|}{CP ($\Delta$) }
	 & 	 & 	\multicolumn{1}{c}{HEFT ($\Delta$) }	& 	\multicolumn{1}{c}{CP ($\Delta$) }	 \\
\hline
Mod 5\_4	 & 	6,328 	 & 	6,076 	(3.98\%) & 	5,984 	(5.44\%) & 	6,016 	 & 	5,670 	(5.75\%) & 	5,578 	(7.28\%)  	 \\
VBE-Adder\_3	 & 	19,126 	 & 	18,972 	(0.81\%) & 	18,852 	(1.43\%) & 	11,416 	 & 	11,148 	(2.35\%) & 	11,148 	(2.35\%)  	 \\
CSLA-MUX\_3	 & 	22,238 	 & 	21,444 	(3.57\%) & 	21,438 	(3.60\%) & 	19,774 	 & 	18,914 	(4.35\%) & 	18,778 	(5.04\%)  	 \\
RC-Adder\_6	 & 	28,606 	 & 	27,584 	(3.57\%) & 	27,564 	(3.64\%) & 	20,408 	 & 	19,136 	(6.23\%) & 	18,906 	(7.36\%)  	 \\
Mod-Red\_{21}	 & 	33,160 	 & 	32,640 	(1.57\%) & 	32,348 	(2.45\%) & 	28,320 	 & 	27,570 	(2.65\%) & 	27,494 	(2.92\%)  	 \\
Mod-Mult\_{55}	 & 	13,942 	 & 	13,822 	(0.86\%) & 	13,740 	(1.45\%) & 	14,478 	 & 	14,270 	(1.44\%) & 	14,070 	(2.82\%)  	 \\
Toff-Barenco\_3	 & 	12,388 	 & 	12,388 	(0.00\%) & 	12,388 	(0.00\%) & 	4,812 	 & 	4,998 	($-$3.87\%) & 	4,812 	(0.00\%)  	 \\
Toff-NC\_3	 & 	9,108 	 & 	9,108 	(0.00\%) & 	9,108 	(0.00\%) & 	3,818 	 & 	3,818 	(0.00\%) & 	3,818 	(0.00\%)  	 \\
Toff-Barenco\_4	 & 	15,992 	 & 	15,786 	(1.29\%) & 	15,582 	(2.56\%) & 	9,006 	 & 	8,678 	(3.64\%) & 	8,678 	(3.64\%)  	 \\
Toff-NC\_4	 & 	12,272 	 & 	11,834 	(3.57\%) & 	11,834 	(3.57\%) & 	8,016 	 & 	7,698 	(3.97\%) & 	7,626 	(4.87\%)  	 \\
Toff-Barenco\_5	 & 	21,600 	 & 	21,388 	(0.98\%) & 	21,378 	(1.03\%) & 	19,424 	 & 	19,220 	(1.05\%) & 	19,138 	(1.47\%)  	 \\
Toff-NC\_5	 & 	13,536 	 & 	13,002 	(3.95\%) & 	12,984 	(4.08\%) & 	9,262 	 & 	8,920 	(3.69\%) & 	8,920 	(3.69\%)  	 \\
Toff-Barenco\_{10}	 & 	90,940 	 & 	89,418 	(1.67\%) & 	89,248 	(1.86\%) & 	59,282 	 & 	57,678 	(2.71\%) & 	57,282 	(3.37\%)  	 \\
Toff-NC\_{10}	 & 	43,156 	 & 	42,600 	(1.29\%) & 	42,546 	(1.41\%) & 	28,720 	 & 	28,494 	(0.79\%) & 	28,322 	(1.39\%)  	 \\
GF($2^4$)-Mult	 & 	33,160 	 & 	32,508 	(1.97\%) & 	32,492 	(2.01\%) & 	33,646 	 & 	33,160 	(1.44\%) & 	33,148 	(1.48\%)  	 \\
GF($2^5$)-Mult	 & 	37,948 	 & 	36,734 	(3.20\%) & 	36,480 	(3.87\%) & 	44,702 	 & 	44,464 	(0.53\%) & 	44,030 	(1.50\%)  	 \\
GF($2^6$)-Mult	 & 	63,856 	 & 	62,174 	(2.63\%) & 	61,870 	(3.11\%) & 	64,784 	 & 	64,278 	(0.78\%) & 	64,180 	(0.93\%)  	 \\
\hline
\end{tabular}
\end{table*}

\appendix
We show how the Heterogeneous-Earliest-Finish-Time (HEFT) algorithm for task scheduling can be used for quantum operation scheduling with a slight modification.
The original HEFT algorithm is designed for scheduling with a soft resource constraint, i.e., every operation can be executed on any processor with a different cost.
We adjust it here so that it can work with a hard resource constraint, i.e., every operation has fixed qubit operands that are not interchangeable.

The original HEFT algorithm consists of two phases:
an \emph{operation prioritizing phase} for computing the priorities of all operations based on upward ranking and
a \emph{processor selection phase} for scheduling the highest priority operation at the moment on the processor, which minimizes the operation's finish time~\cite{topcuoglu2002performance}.
In the processor selection phase, the algorithm considers the possibility of inserting an operation in the earliest idle time-slot between two already scheduled operations.
Only the idle time-slots that preserve precedence constraints, i.e.,
that comply with the dependency graph, are considered in this phase.
This \emph{insertion-based policy} allowing the insertion in the idle time-slot characterizes  the HEFT algorithm.

While keeping this insertion-based policy, we adjust the HEFT algorithm so that every operation is assigned to the fixed qubits (i.e., processors in the original term), which means we no longer need to select qubits in the processor selection phase.
Note that it is necessary for the adjusted HEFT algorithm to maintain scheduled time-slots across qubits,
whereas the original algorithm simply maintains the time-slots by processors.
The process flow of the HEFT algorithm for quantum operation scheduling is shown in Algorithm \ref{algo:heuristic}.

\begin{algorithm}[htb]
		\small
	\caption{HEFT algorithm for QOS}
	\label{algo:heuristic}
	\begin{algorithmic}[1]
		\State $G = (V, C)$: dependency graph of a QOS problem
		\State Compute upward rank $r(u)$ for each operation $u \in V$ by
				$$ r(u) = d(u) + \max_{v \in succ(u)}{r(v)} $$
				where $succ(u)$ is the set of immediate successors of $u$,
				$d(u)$ is the duration of $u$, and $r(e) = d(e)$ for any exit operation $e$.
		\State $ready\_time(u) = 0$ for all $u \in V$.
		\For{all $u \in V$ in descending order of  $r(u)$}
			\State Insert $u$ at the start time $t$ of the earliest idle time-slot (whose duration $> d(u)$) after $ready\_time(u)$.
			\For{all $v \in succ(u)$}  
				\State $ready\_time(v) = \max(ready\_time(v), t + d(u))$.
			\EndFor
		\EndFor
	\end{algorithmic}
\end{algorithm}

We conducted experiments to check the solution quality of the adjusted HEFT algorithm with the same benchmark sets and experimental settings as used in Section~\ref{sec:experiment}.
For all of the instances under the formulation with extended DAG (Ext-DAG),
the HEFT algorithm always succeeded in finding solutions slightly worse than or equal to those by the CP solver.
The medians of improvement rates from Std-DAG to Ext-DAG with the HEFT algorithm (the CP solver) were
1.16\% (1.26\%) for circuits by Zulehner et al.~\cite{zulehner2018efficient} used in Table~\ref{tab:improvement} and
1.67\% (2.45\%) and 2.35\% (2.82\%) for circuits by Nam et al.~\cite{nam2018automated} used in Table~\ref{tab:diff} with naive and optimized gate decomposition, respectively.
This suggests that the HEFT algorithm is a good option for adding a bit more optimization in cases where not much compilation time is available.

All the results of these latter two experiments are listed in Table~\ref{tab:heft}.
We can see a negative improvement at Toff-arenco\_3 using optimized gate decomposition and the HEFT algorithm.
This can happen because considering further commutation in the formulation with Ext-DAG yields a broader search space for algorithms,
and it provides an opportunity to find not only a better solution than Std-DAG but also a worse one.
However, this is not a big issue in practice in cases where we can afford to select the better of the solution by the as-soon-as-possible algorithm with Std-DAG and that by the HEFT algorithm with Ext-DAG.

\end{document}